\documentclass[%
reprint,
superscriptaddress,
 amsmath,amssymb,
 aps,
prl,
floatfix,
]{revtex4-1}

\usepackage{graphicx}
\usepackage{dcolumn}
\usepackage{bm}
\usepackage[mathlines]{lineno}

\usepackage{siunitx}

\usepackage{xspace}
\newcommand{\bise} {\ensuremath{{\mathrm{Bi}_2\mathrm{Se}_3}}\xspace}

\newcommand{\bite} {\ensuremath{{\mathrm{Bi}_2\mathrm{Te}_3}}\xspace}

\newcommand{\cpara} {\ensuremath{{c\parallel B_0}}\xspace}
\newcommand{\cperp} {\ensuremath{{c\bot B_0}}\xspace}

\newcommand{\ba}{\textsf{A}\xspace}
\newcommand{\bb}{\textsf{B}\xspace}
\newcommand{\bc}{\textsf{C}\xspace}
\newcommand{\bd}{\textsf{D}\xspace}
\newcommand{\be}{\textsf{E}\xspace}
\newcommand{\bff}{\textsf{F}\xspace}

\begin{document}

\preprint{APS/123-QED}

\title{Field induced charge symmetry in topological insulator Bi$_2$Te$_3$\\ revealed by nuclear magnetic resonance}

\author{R. Guehne}
\email{r.guehne@physik.uni-leipzig.de}
\author{J. Haase}%
 
\affiliation{%
 Felix Bloch Institute for Solid State Physics,
 Leipzig University, Linn\'estraße 5, 04103 Leipzig, Germany
}%

\author{C. Shekhar}
\author{C. Felser}
\affiliation{
 Max Planck Institute for Chemical Physics of Solids,
 N\"othnitzer Straße 40, 01187 Dresden, Germany 
}%


\date{\today}

\begin{abstract}
Nuclear magnetic resonance (NMR) was recently shown to measure the bulk band inversion of Bi$_2$Se$_3$ through changes in the $^{209}$Bi nuclear quadrupole interaction, and the corresponding tensor of the local electric field gradient was found to follow, surprisingly, the direction of the external magnetic field if the sample is rotated. This manifests a hidden property of the charge carriers in the bulk of this topological insulator, which is explored here with another material, Bi$_2$Te$_3$. It is found that two electric field gradients appear to be present at $^{209}$Bi, one rests with the lattice, as usual, while a second follows the external field if it is rotated with respect to the crystal axes. These electronic degrees of freedom correspond to an effective rotation of $j$-electrons, and their level life time is believed to be responsible for a new quadrupolar relaxation that should lead to other special properties including the electronic specific heat.
\end{abstract}


\maketitle

\noindent
Gapless surface states of \bise-type 3-dimensional topological insulators are of great interest from fundamental as well as applied physics' perspective \cite{Ando2013}, and they have been studied extensively with surface sensitive methods \cite{Xia2009,Zhang2009a,Zhang2012}. A spin-orbit coupling (SOC) induced energy band inversion at the $\Gamma$-point in the Brillouin zone of these materials was anticipated to lead to these topologically protected states \cite{Zhang2009}. However, related bulk properties of these materials have hardly been investigated.

Nuclear magnetic resonance (NMR) as a local, bulk probe was used, in a few applications, to study magnetic properties, in particular with aim at the surface states  by using nano-powders, to increase the surface area \cite{Koumoulis2013,Papawassiliou2020}. However, definite conclusions are difficult as the NMR of spin-orbit coupled systems is not well understood \cite{Boutin2016}, and bulk properties are expected to be altered by very small grains, as well \cite{Gioia2019}. For example, a special Bloembergen-Rowland type electronic susceptibility was found to be present in these narrow gap materials as it leads to large, field independent NMR line widths \cite{Georgieva2016}.

Very recently, some of us have shown that NMR can indeed detect the band inversion of itinerant carriers via its effect on the local charge symmetry that influences the nuclear levels through the electric quadrupole interaction \cite{Guehne2019}, leading to the well known splitting of NMR lines for spin $I>1/2$ nuclei \cite{Slichter1990}. This quadrupole splitting is known to be very sensitive to the local charge symmetry represented by the electric field gradient (EFG) at the nuclear site $-$ it even detects lattice strain arising from impurities. It is thus not surprising that quadrupole splittings also reflect changes of the electronic wave function due to SOC. Furthermore, the experiments can be reliably compared to first-principle calculations. 

\begin{figure}[t]
\includegraphics[width=\columnwidth]{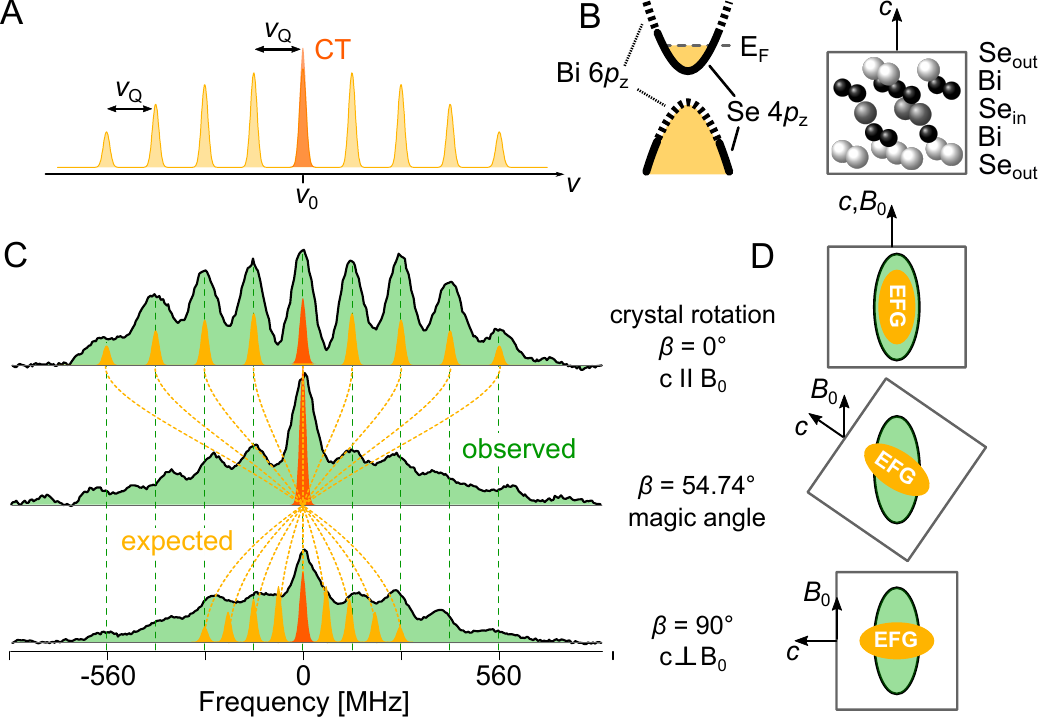}
\caption{\label{fig:0} $^{209}$Bi NMR of single crystalline \bise: \ba, expected quadrupole splitting, $\nu_Q$, in the high magnetic field $B_0$ case. The tensor of the electric field gradient (EFG) at the Bi nucleus and its orientation with respect to the external field determine $\nu_{\mathrm Q}$. \bb, the crystal symmetry at the Bi site \emph{and} the band structure (band topology) affect $\nu_{\mathrm Q}$ through local changes in the wave function or carrier concentration \cite{Guehne2019}. \bc, comparison with measurements (green spectra) are in quantitative agreement with DFT calculations \cite{Guehne2019}, however, the expected orientation dependence of $(3\cos^2\beta-1)$ from this second rank tensor (orange spectra) is lacking, on the contrary, an orientation \emph{in}dependent splitting (green spectra and lines) suggests that the EFG rotates with $B_0$, rather than being tied to the crystal, as depicted in \bd.}
\end{figure}

Very unusual, however, and unfortunately not easily tractable with first-principle calculations, is the orientation dependence of the quadrupole splitting in \bise \cite{Guehne2019}, i.e., if one changes the orientation of the crystal with respect the external magnetic field ($B_0$). In NMR it is always assumed that the EFG, a second-rank tensor, rests with the crystal axes system. For example, this leads to typical powder spectra when the grains are oriented randomly, or, in case of single crystals, to special orientation dependences of the NMR spectra, which are a fingerprint of the local crystal symmetry. Surprisingly, for \bise, this tensor appears to follow the external field to a large extent, as shown in~Fig.\ref{fig:0}.

In leading order of the nuclear Zeeman interaction, the electric quadrupole splitting is usually expressed in terms of an orientation dependent quadrupole frequency ($\nu_{\mathrm Q}$) given by the orientation dependent influence of the  second rank tensor of the local electric field gradient ($V_{\mathrm{XX}}, V_{\mathrm{YY}}, V_{\mathrm{ZZ}}$). The local symmetry at the Bi nucleus requires that $V_{\mathrm{XX}}=V_{\mathrm{YY}}$ and only one tensor component enters the well-known expression $\nu_Q  = {3eQV_{\mathrm{ZZ}}}/({{4I\left( {2I - 1} \right)h}})\cdot({{3\cos^2\beta  - 1}})$,
where $I$ is the nuclear spin and $eQ$ the electric nuclear quadrupole moment. The azimuthal angle $\beta$ spans between $V_{\mathrm{ZZ}}$ and the external magnetic field $B_0$. 

For $^{209}$Bi in \bise, $V_{\mathrm {ZZ}}$ is strongly influenced by the band inversion and depends on the carrier concentration \cite{Guehne2019}. However, there is hardly any orientational dependence, i.e., $V_{\mathrm {ZZ}}$ remains nearly aligned with $B_0$, cf.~Fig.\ref{fig:0}~\bc. Interestingly, this new phenomenon is most prominent for higher carrier concentrations, $n\geq\SI{E19}{cm^{-3}}$ \cite{Guehne2019}, so that one has to associate it with the doped electrons. It was argued that the system's massive SOC ($g_{\mathrm{eff}}\approx 30$ \cite{Kohler1975}) allows the external magnetic field to deform partially filled Bi orbitals. The resulting EFG is then set by $B_0$, in stark contrast to an ordinary EFG that is pinned to the crystal axes. With other words, the effect involves rotational degrees of freedom of strongly spin-orbit coupled conduction electrons, with their orbital angular momenta being partially unquenched due to $B_0$ \cite{Sushkov2019}.
Earlier $^{209}$Bi NMR studies of \bise \cite{Young2012,Nisson2013,Nisson2014,Mukhopadhyay2015} are in agreement with the above mentioned special results, but the short nuclear relaxation times for Bi apparently hindered a clear identification of this new effect, before \cite{Nisson2014}.

In order to shed more light on this unusual behavior, we decided to investigate \bite, the closest relative of \bise, with $^{209}$Bi NMR. We find the quadrupole splitting to be an order of magnitude larger compared to that in \bise, and the orientation dependent splitting appears closer to what is expected form \textit{ordinary} behavior. However, near the magic angle ($\cos^2\beta = 1/3$), where the usual quadrupole splitting disappears, it becomes apparent that a quadrupolar split spectrum remains, similar to what was observed in \bise, manifesting a clear discrepancy with the expected behavior. In fact, a closer look at the total angular dependence shows this phenomenon also at other angles. Again, it appears that the electronic degrees of freedom change with the orientation of the external field, yielding a non-vanishing splitting even at the magic angle. In contrast to \bise, however, the quadrupole interaction in \bite appears to have two EFGs,  a large lattice component that rotates together with the sample, and a second component that follows the field. Hence, the $^{209}$Bi quadrupole interaction in \bite provides a new perspective on a very unusual effect related to so far unknown electronic properties of conduction electrons in topological insulators.

\bite used for the investigations here were grown by self-flux method. A stoichiometry ratio of Bi (purity: \SI{99.999}{\%}) and Te (purity: \SI{99.999}{\%}) are sealed in a dry quartz ampoule with vacuum of \SI{E-5}{mbar}. Except for the cooling rate of \SI{2}{K/h} from \SI{800}{} $-$ \SI{500}{\degreeCelsius} for the crystal growth, the other heating profile is the same as in ref. \cite{ViolBarbosa2013}. A c-axis oriented platelet-like crystal of several mm in dimension was mechanically separated from the ingot.

All experiments were carried out at room temperature with commercial NMR consoles using a home-built probe that fits standard \SI{7.05}{T}, \SI{11.74}{T}, and \SI{17.6}{T} magnets. We employed spin echoes ($\pi/2-\tau-\pi$) to measure individual transitions, their decay ($T_2)$, as well as their recovery after saturation ($T_1$), besides, solid echoes ($\pi/2-\tau-\pi/2$) to excite the whole set of transitions simultaneously where possible.

\begin{figure}[t]
\includegraphics[width=\columnwidth]{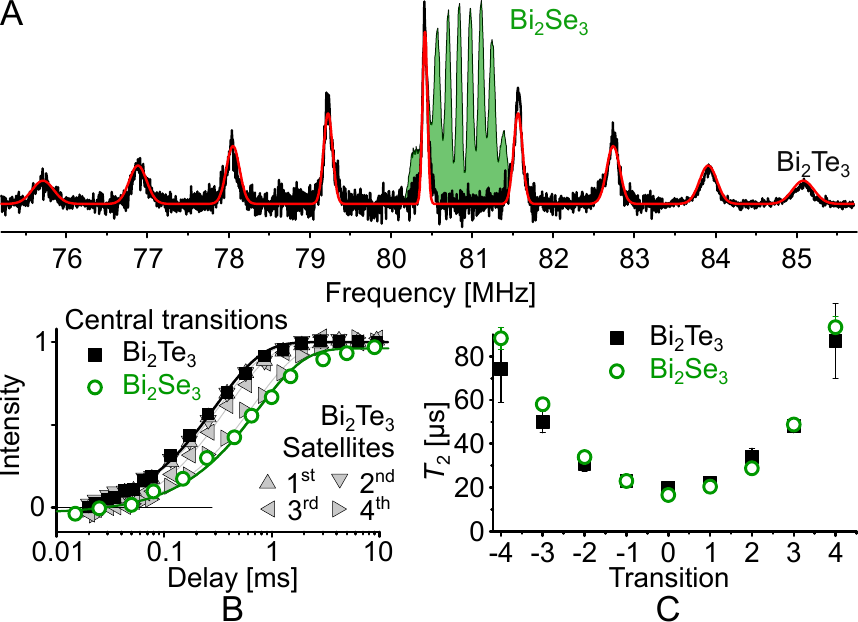}
\caption{\label{fig:1} \ba, quadrupole spectrum of $^{209}$Bi in \bite (black) obtained for \cpara at \SI{11.74}{T} (selective spin echoes with $\pi/2$-pulse width of \SI{3}{\mu s}), including a fit (red line, see text for details). Also shown is a typical $^{209}$Bi spectrum of single crystalline \bise (S3 in \cite{Guehne2019}). \bb, saturation recovery of the Bi central transition in \bite (squares) and \bise (circles), respectively, next to the satellite recoveries  in \bite (triangles). \bc, transition selective spin-echo decay times, $T_2$, in \bite and \bise.}
\end{figure}

We start with $^{209}$Bi NMR spectra of the \bite single crystal with the magnetic field $B_0$ along the crystal $c$-axis (\cpara) where $\beta=0$, cf.~Fig.~\ref{fig:1}\ba.

We find the expected $2I=9$ resonances and a simple fit of Gaussian lines to the total spectrum is in very good agreement with the expected pattern for a magnetic (the same for all transitions) and quadrupolar (contribution changes with transitions and is absent for the central line) broadening; we find $\nu_Q=\SI{1.171(2)}{MHz}$, a central transition linewidth of \SI{80(2)}{kHz}, and a relative quadrupole broadening of $\SI{7}{\%}$, i.e. about $\SI{80}{kHz}$ for the first pair of satellites. For comparison, we show the spectrum for \bise, as well (sample S3 in \cite{Guehne2019}). The quadrupole splitting frequency is \SI{0.14}{MHz} with a central transition width of \SI{77(4)}{kHz} and a relative quadrupolar boradening of up to \SI{9}{\%}, i.e. $\sim\SI{13}{kHz}$ for the first satellites \cite{Guehne2019}. Thus, while the quadrupole coupling and its distribution are nearly a factor of 10 larger in \bite, the magnetic (central) widths are similar. This NMR line broadening agrees with theoretical linewidths of an enhanced nuclear dipole coupling arising from the Bleombergen-Rowland electronic spin susceptibility \cite{Georgieva2016,Jorge2017}.

Comparison of the relaxation in \bite and \bise reveals similarities, as well. Selective recoveries of the central transitions are shown in Fig.\ref{fig:1}~\bb, with apparent recovery times of \SI{303(5)}{\mu s} for \bite and \SI{750(50)}{\mu s} for \bise. Since individual transitions are well separated in the case of \bite, we measured their recovery, as well (triangles). The recovery slows down with increasing satellite order, from \SI{311(5)}{\mu s} (1$^{\mathrm{st}}$ satellite) over \SI{337(5)}{\mu s} and \SI{382(5)}{\mu s} (2$^{\mathrm{nd}}$ and 3$^{\mathrm{rd}}$, respectively) to \SI{672(10)}{\mu s} (4$^{\mathrm{th}}$). Comparison with a purely magnetic relaxation \cite{Takigawa1991,Nisson2013} based on the recovery of the central transition reveals discrepancies, especially for the outer satellites, indicating contributions form quadrupolar relaxation \cite{Kranendonk1954,Suter1998}.

Selective spin-spin relaxation experiments, Fig.\ref{fig:1}~\bc, yield the same $T_2$ values within error for both samples, with a characteristic increase in $T_2$ with growing satellite order. The central transition $T_2$ values are further in good correlation with their width in frequency domain, and thus with the aforementioned model calculations based on the special Bloembergen-Rowland type of indirect nuclear dipole coupling.

\begin{figure}[!ht]
\centering
\includegraphics[width=.98\columnwidth]{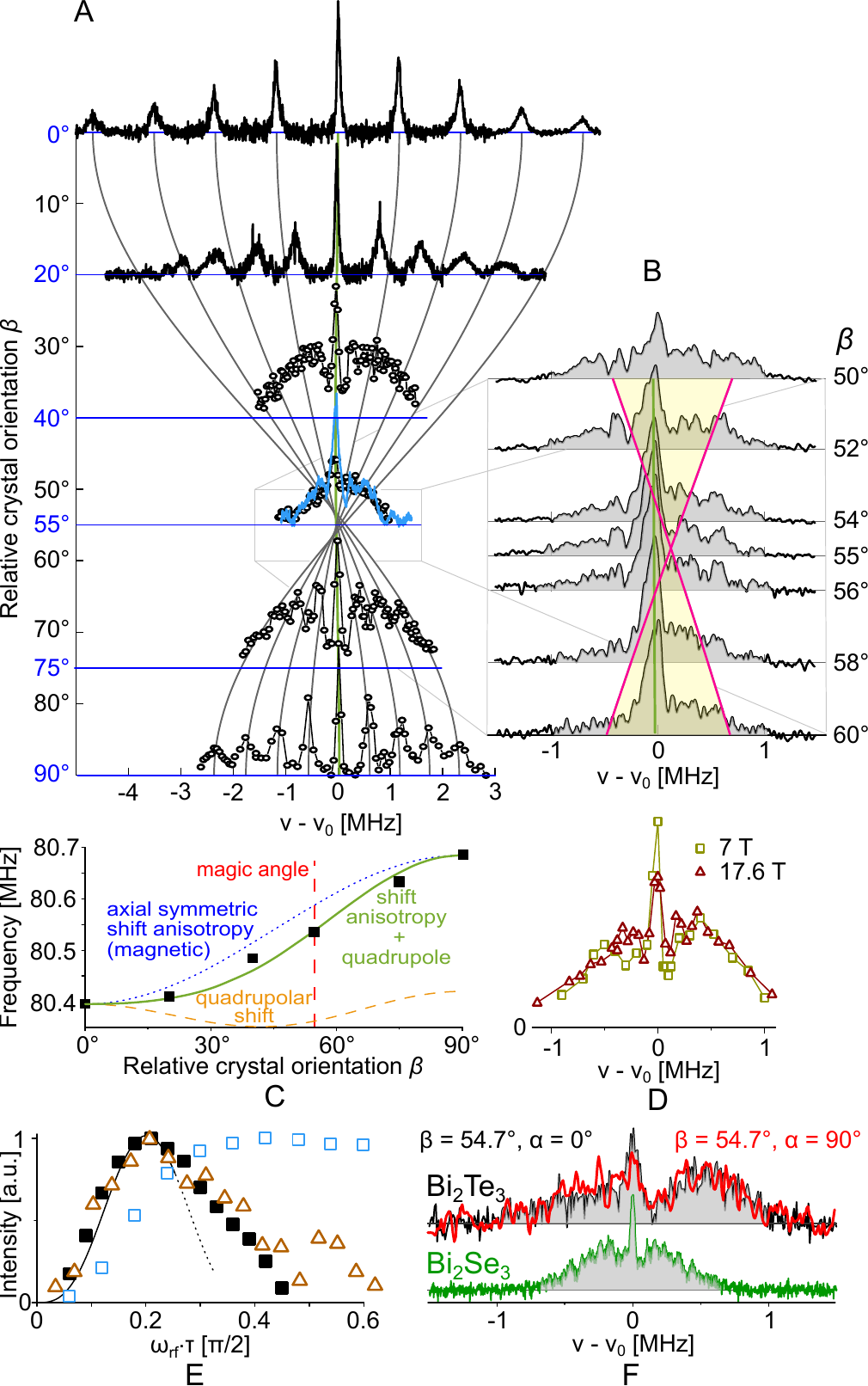}
\caption{\label{fig:2} Orientation dependent $^{209}$Bi NMR spectra of \bite (black spectra represent individually measured transitions, black circles frequency swept spectra, and the blue spectrum was obtained from a solid-echo experiment). Main panel \ba, the top trace for \cpara ($\beta=\ang{0}$) is connected with what is expected from the quadrupole interaction by grey lines with the bottom trace \cperp ($\beta=\ang{90}$).  
 The inset to the right, \bb, shows a magnified window of solid-echo experiments between $\beta=\ang{50}$ and \ang{60}. The 2 pink lines give the traces of the two outermost satellites, i.e. spectral density is only in the yellow area. In contrast, the measured spectra cover a much larger range of about $\pm\SI{1}{MHz}$ (grey filling). \bc, the resonance frequency of the central transition as a function of $\beta$ reflects an axial symmetric Knight shift anisotropy plus higher order quadrupole interaction (solid line).  \bd, frequency swept spin echoes measured at $B_0=\SI{7}{T}$ (squares) and \SI{17.6}{T} (triangles) for the magic angle. \be, selective spin-echo nutation for the magic angle $+$\SI{240}{kHz} off center (triangles) in agreement with the CT nutation for \cpara (solid squares, line) confirms a quadrupolar origin of the broad magic angle spectrum. Consequently, broad band solid echo nutation (open squares) behave differently, with a much slower nutation frequency. \bff, magic angle broad band solid echo spectra for $\tau=\SI{32}{\mu s}$ of \bite ($\alpha=\ang{0}$ and \ang{90}) and \bise.}
\end{figure}

We now turn to orientation dependent spectra shown in Fig. \ref{fig:2}~\ba. The gray solid lines represent the expected resonance frequencies as a function of $\beta$, as obtained from exact diagonalization of the total Hamiltonian with $\nu_{\mathrm{Q}}=$\SI{1.171}{MHz} for an axially symmetric field gradient, at a resonance frequency of \SI{80.4}{MHz} ($B_0=\SI{11.74}{T}$), since higher order effects have to be included.  While the central transition follows the predicted frequency dependence on angle (Fig.\ref{fig:2}~\bc), the satellite resonances do not. In particular, near the magic angle the total satellite spectral weight is distributed over a far too large a spectral range, while some asymmetry of the spectrum is expected from higher order effects. The broad spectral range and its asymmetry follow from broadband solid echo spectra as well as from frequency stepped, selective excitation. A more detailed orientation dependence near the magic angle is shown in panel~\bb, with the expected (yellow area) and the measured (gray areas) spectral intensity (green line is the expectation for the central transition). With the magnetic field perpendicular to the $c$-axis the spectrum approaches what is expected from \cpara. The question arises how one can understand the disagreement, in particular near the magic angle.

Frequency stepped spin-echo spectra near the magic angle show, cf. Fig.~\ref{fig:2}\bd, that the spectrum is independent on the field, ruling out higher order effects from a much larger quadrupole term that could perhaps appear, but would also not fit the central transition position easily (only with assumptions about a special magnetic shift). Selective nutation (panel~\be) reveals that the signal intensity belongs to single transitions, so it must be given by quadrupolar satellites that extend over a range of about 1 MHz. Note that the total spectral intensity remains constant as a function of $\beta$, thus, the observed effects concern the bulk signal. This means, the spectral intensity near the magic angle cannot be accounted for by sample inhomogeneity. 

Obviously, the description above points immediately to a second quadrupolar broadening mechanism (of about \SI{500}{kHz}) as it produces the desired broadening, i.e., the splitting up of the satellite transitions without affecting the central transition and keeping the asymmetry (center of gravity is unchanged). However, this additional broadening must be rather independent from the main broadening mechanism and, given the lack of a special lineshape, is likely to carry an additional inhomogeneity. Since the orientation dependence of the spectra is not very good at other angles, it is likely that the second term is already present far away from the magic angle, however, in view of the rather narrow lines for \cpara, we believe that it must be significantly smaller for this direction of the field, and possibly also for \cperp. 

Note, that symmetry in the plane is conserved as a rotation about the crystal $c$-axis ($\alpha=\ang{0},\ \ang{90}$) inside the NMR coil does not change the spectrum, cf.~Fig.~\ref{fig:2}~\bff, as expected.

Given the similarity of a non-vanishing magic angle quadrupole splitting in both materials, \bite and \bise, cf.~Fig.~\ref{fig:2}\bd, and their close electronic relationship, i.e., very large effective $g$-factors of conduction electrons \cite{Kohler1975,Kohler1976}, we argue that quadrupole coupling that follows the field arises from the effect of the magnetic field on electrons in particular $j$-states, while the much stronger (normal) Bi quadrupole interaction in \bite is related to a large EFG component from the ionic lattice. Note that for \bise the effect of a non-trivial, orientation independent quadrupole splitting due to $j$-electrons dominates the observation, with a negligible lattice EFG ($\nu_Q^{\mathrm{latt}}\leq\SI{10}{kHz}$). In \bite, the opposite situation holds, with a dominating lattice EFG yielding a first order orientation dependence (with small 2$^{\mathrm{nd}}$ order corrections), which gives way to a non-trivial EFG component from $j$-electrons (precessing about $B_0$) dominating the NMR near the magic angle. 

In the last part of the discussion, we reconsider spin-lattice relaxation in the context of a special EFG arising from these electrons final life time of precession, as it must result in fluctuations of the EFG and thus nuclear relaxation. Results on the recovery of individually saturated Bi transitions in \bite for \cpara indicate a quadrupolar spin-lattice relaxation. The estimated recovery times of the central transitions in \bite and \bise, however, are in clear disagreement with the measured quadrupole splittings for \cpara, because quadrupolar relaxation is expected to be proportional to $\nu_Q^2$. One may thus invoke contributions from magnetic relaxation. Though, when comparing the magic angle spectra for \bite and \bise in Fig.\ref{fig:2}~\bff, both patterns are rather similar, the one obtained from \bite is only slightly larger then the one from \bise. Hence, an alternative scenario to account for Bi spin-lattice relaxation  in \bite and \bise could be related to the non-trivial EFG component of $j$-electrons that may be related to a special electronic thermal reservoir.

In summary, $^{209}$Bi NMR of \bite reveals a non-trivial quadrupole interaction in the sense that the local charge distribution is altered by an external magnetic field. This bears strong similarities to what was observed in \bise, an electronically similar material. This behavior points to low energy electronic degrees of freedom in \bite and \bise. Not only is this behavior crucial for the understanding and analysis of NMR data in such materials, it also raises the question of the role of these excitations for other material properties, for example their so far unconsidered potential to contribute to the excellent thermoelectric performance \cite{Fu2020}.

RG and JH thank O. Sushkov, I. Garate, G. Williams, S. Chong, N. Georgieva, J. Nachtigal, A. Isaeva, and O. Oeckler for stimulating discussions. We acknowledge the financial support by the Deutsche Forschungsgemeinschaft, project 442459148, and by Leipzig University. The samples were provided for this study under the ERC Advanced Grant No. 742068 TOPMAT.

\bibliography{references}

\end{document}